# The PRISMA-36 array for studying variations of the thermal neutron flux

M.B. Amelchakov[1], A. Chiavassa[2,3*], D.M. Gromushkin[1], S.S. Khokhlov[1], E.P. Khomchuk[1], V.V. Kindin[1], A.Yu. Konovalova[1], P.S. Kuzmenkova[1], E.S. Morgunov[1], N.A. Pasyuk[1], A.A. Petrukhin[1], I.A. Shulzhenko[1], E.P. Volkov[1], I.I. Yashin[1]

[1] National Research Nuclear University MEPhI (Moscow Engineering Physics Institute), 115409, Moscow, Russia
[2] Dipartimento di Fisica Dell' Universit`a Degli Studi di Torino, 10125, Torino, Italy
[3] Sezione di Torino Dell' Istituto Nazionale di Fisica Nucleare – INFN, 10125, Torino, Italy

*Corresponding author E-mail address: andrea.chiavassa@unito.it

### Abstract

From 2012 to 2023, the PRISMA-32 array was in operation at the Experimental Complex NEVOD (MEPhI, Moscow). The purpose of the array was to study extensive air showers by detecting the air-shower neutron and electron-photon components using unshielded neutron detectors. To expand the capabilities of this facility, including for the study of cosmic and geophysical phenomena with a neutron flux, its upgrade was carried out. During the upgrade, a dedicated measuring channel for studying variations of the neutron background and the processes affecting these variations was created. To achieve this, the photomultipliers, the integrating amplifiers, the digitalizing electronics and the high-voltage power supply system were replaced. The paper describes the structure of the upgraded array, which was named PRISMA-36, and presents the results of studying the characteristics of the main elements of its "variation" channel. A method for identifying signals caused by neutron capture and the determined criteria for their selection are discussed. An example of a Forbush decrease, caused by a X1.1-class flare and recorded with the variation channel of the PRISMA-36 array, is given.

### 1. Introduction

The study of variations in the intensity of cosmic ray (CR) flux is a separate section of Cosmic Ray Physics. Research in this area allows obtaining information about both the state of the heliosphere and the physical processes and phenomena on the Earth [1, 2]. Today, the continuous observations of cosmic ray variations are performed by a network of neutron monitors installed on the ground surface and recording the hadron component of CR in the primary energy range of 1-100 GeV, as well as by the muon telescopes and hodoscopes sensitive to primary CR with energies up to a 1 TeV. The studies conducted at these facilities can be divided into several main areas: long-term [3] and short-period [4] CR variations, cosmic ray anisotropy [5], solar CR [6, 7], atmospheric effects [8], etc.

In recent years, the thermal neutron detectors based on ZnS(Ag)+6LiF [9, 10] or ZnS(Ag)+B2O3 [11, 12] scintillators have become widely used. The use of such detectors in an unshielded form makes it possible to study variations of the neutron flux that is in thermal equilibrium with the environment. This type of detector can be used both to study the cosmic ray variations and the environment itself, as well as the various geophysical processes (tidal waves of the Earth's crust, natural oscillations of the Earth, etc.). A feature of such detectors is the opportunity to select signals caused by the capture of neutrons by $^6$Li and $^{10}$B nuclei (hereinafter referred to as "neutron signals") based on the pulse shape. This selection makes it possible to minimize the contribution of photodetector noise and external interference, as well as of charged particles. Using similar detectors, a number of studies have been carried out, and the first results have been obtained on: the Forbush decreases [13, 14, 15], the detection of neutrons during thunderstorms [16], the influence of meteorological parameters on the neutron flux variations [17], the search for relationship between variations and different geophysical phenomena [18, 19].

The unshielded thermal neutron detectors with the ZnS(Ag)+$^6$LiF scintillator were the basis of the PRISMA-32 array (MEPhI, Moscow) for measuring the electron-photon and neutron components of extensive air showers (EAS) [9]. The PRISMA-32 array was launched into operation in 2012 and provided the first results on EAS neutrons. It became the prototype for the URAN [20], ENDA-INR [21] and ENDA-LHAASO [10] facilities. In 2023–2024 the array was upgraded in order to expand the range of implemented tasks. As a result, the number of detectors was increased to 36 (total scintillator area is about 13 m$^2$), their layout was optimized, and the photomultipliers and measuring electronics were replaced. In addition to EAS research, the array's tasks now include studying variations of the neutron flux near the Earth's surface using an additional variation channel that has been implemented. The upgraded PRISMA-36 facility will make it possible to study variations of thermal neutrons associated with both the changes in the primary CR intensity and with the processes of geophysical origin.

A large-sized hemispherical photomultiplier EMI 9350KA is used as a new photodetector for the array. This type of PMT has not been previously used in similar detectors. Therefore, the paper pays special attention to the study of the PMT characteristics, to the spectrometric path of the measuring channel, as well as to the method for selecting signals caused by the capture of thermal neutrons.

## 2. The PRISMA-36 array

### 2.1. Array layout

The PRISMA-36 array consists of 36 detectors, combined into three independent clusters of 12 detectors in each. The detectors are deployed in a temperature-stabilized hall (24±1 °C) on the third floor of the Experimental Complex (EC) NEVOD building. The layout of the PRISMA-36 array detectors is shown in Fig. 1. The typical distances between detectors are 3.6 m and 5.3 m along the long and the short sides of the building, respectively. The total area of the array is about 500 m$^2$. Some deviations from the regular grid (white circles) are due to the location of other facilities, technological systems and communications of the EC NEVOD. The detectors are connected to the local post (LP) for data collecting and primary processing of the PRISMA-36 using the cable lines of the same length.

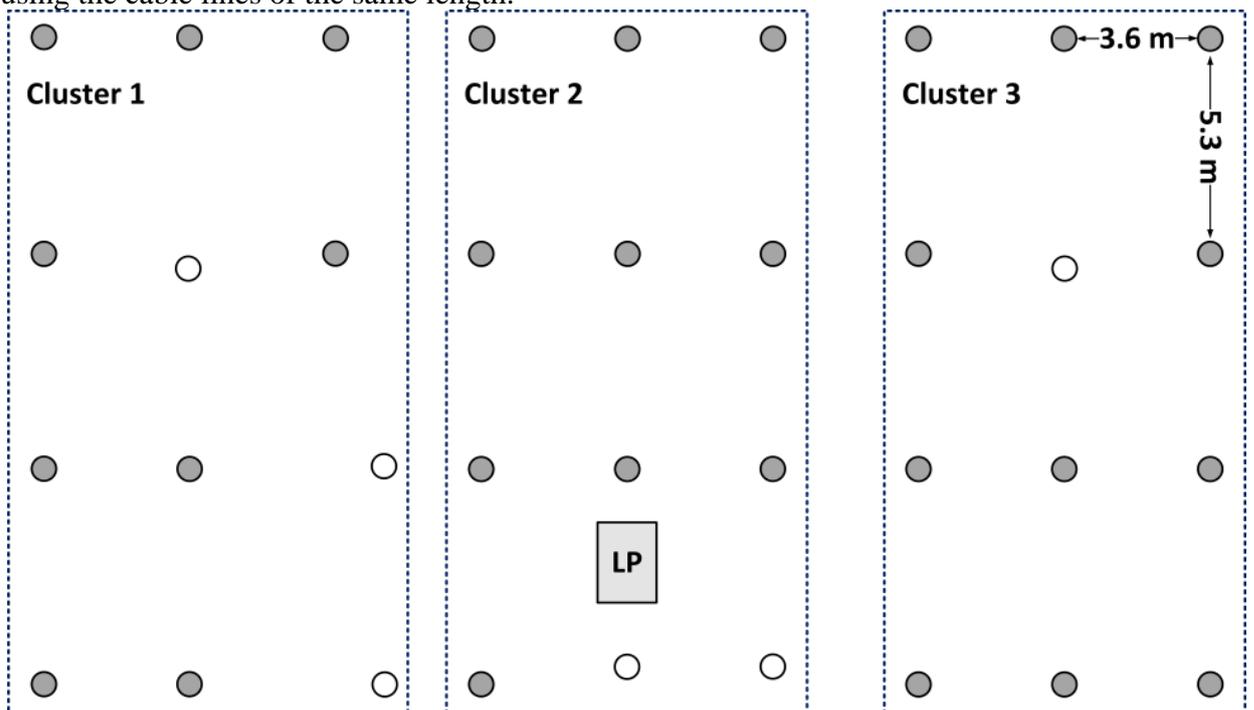

*Fig. 1. Layout of the PRISMA-36 array detectors (circles).*

The main elements of the PRISMA-36 array are:
- the scintillation detectors for recording thermal neutrons;
- the measuring system for digitalizing signals from scintillation detectors and for channels' synchronizing.

### 2.2. Scintillation detector

The PRISMA-36 neutron detector is a light-proof housing (height of 570 mm, 740 mm in diameter), on the bottom of which a thin layer (~ 30 mg/cm$^2$) of inorganic scintillator SL6-5 (ZnS(Ag) + LiF, Li is enriched to 90%) is installed. The scintillator is viewed by a single photomultiplier tube (PMT), which, together with a divider-integrator-amplifier (DIA), is attached to a removable lid of the detector housing. To improve light collection, a light-collecting cone made of aluminium foil cover polyethylene foam with a high diffuse reflection coefficient (~ 90%) is installed inside the housing. The effective area of the scintillator is about 0.36 m$^2$. The detector diagram is shown in Fig. 2.

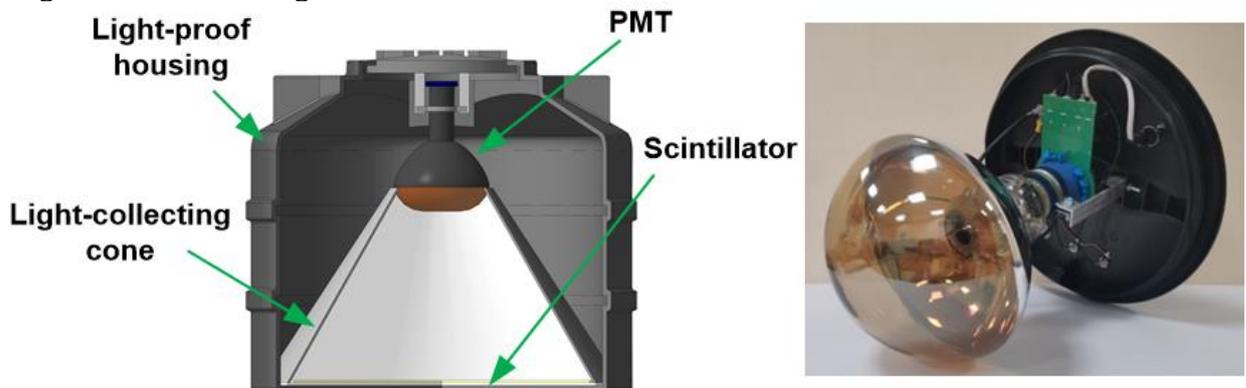

*Fig. 2. The construction of the neutron detector (left) and the photograph of the PMT EMI with the divider-integrator-amplifier (right).*

Thermal neutrons are detected as follows. The neutron is captured by $^6$Li atoms according to the reaction $^6Li + n \rightarrow T + \alpha + 4.78$ MeV. The reaction produces an α-particle and a tritium nucleus, which cause zinc sulfide to emit blue light. When a neutron is captured, up to 160000 photons can be produced. However, due to the low intrinsic transparency of ZnS, the number of photons reaching the scintillator surface is greatly reduced and depends on the depth of interaction within the ZnS granule. That makes neutron detection difficult. A distinctive feature and advantage of ZnS(Ag) is its long scintillation decay time when detecting heavy charged particles [22], which allows rejection of signals caused by neutron capture from PMT noise, external interference, light charged particles and gammas.

### 2.3. Measuring system

The main structural elements of the PRISMA-36 measuring system are the cluster local post (LP) for data collection and primary processing, as well as the central post (CP) for control, synchronization and data acquisition.

The PRISMA-36 local post ensures operation of the PMTs and DIAs of a cluster, digitizes analog signals and selects events according to the specified intra-cluster trigger conditions, performs timestamping of events with an accuracy of 10 ns and transmits information about events to the central post. Functional diagram of the PRISMA-36 local post is shown in Fig. 3.

The cluster local post includes:
- three 12-channel blocks of amplitude analysis of a cluster: BAAC12-100M for detecting neutrons and studying their variations, BAAC12-200T and BAAC12-200M for detecting and studying extensive air showers [20]. The BAAK12 blocks (manufactured by High Technologies LLC, Moscow) are based on FPGA Xilinx Spartan-6 and are installed in a 6U Euromechanics crate;

- the remotely controlled 12-channel high-voltage power supply BVP-12K (manufactured by "Mantigora", Novosibirsk) providing high voltage for detectors' PMTs;
- the power supply providing low voltage (±12 V) for the detectors' DIAs.

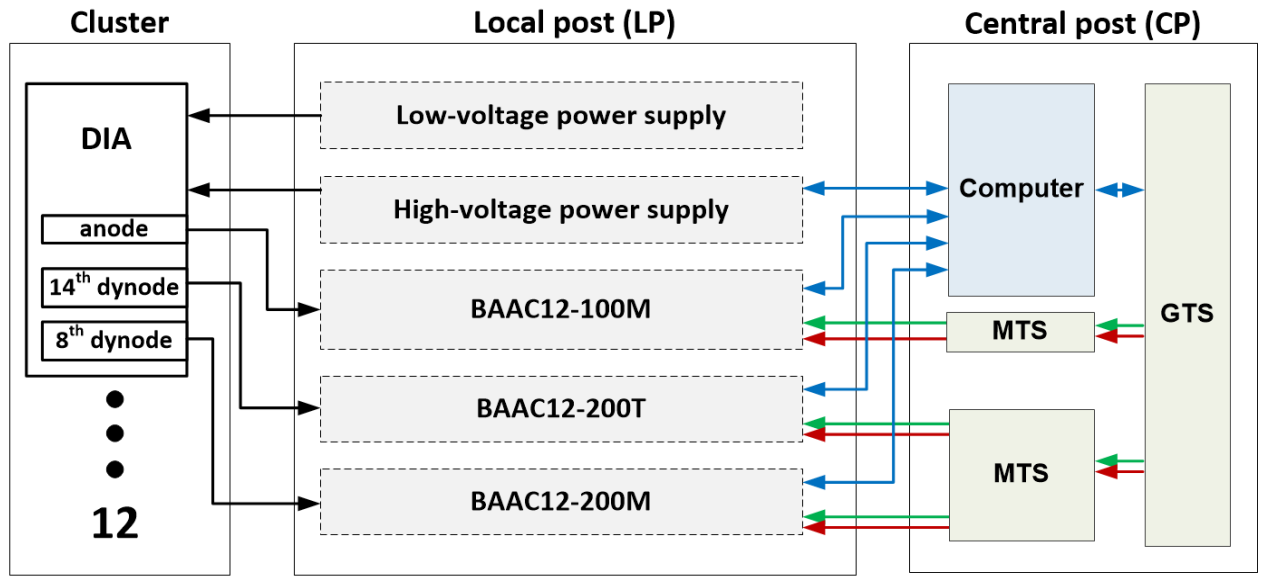

Fig. 3. Functional diagram of the PRISMA-36 local post.

Local posts of three PRISMA-36 clusters are installed inside a single 19-inch telecommunication rack. The LP equipment is connected to the array central post via fiber-optic communication lines.

The central post of the array is located in the Control Room of the EC NEVOD and provides control, time synchronization of BAAC12, as well as their data collection and processing. Its main elements are the control computer of the array, as well as two modules of time synchronization (MTS). Functional diagram of the PRISMA-36 central post is shown in Fig. 4.

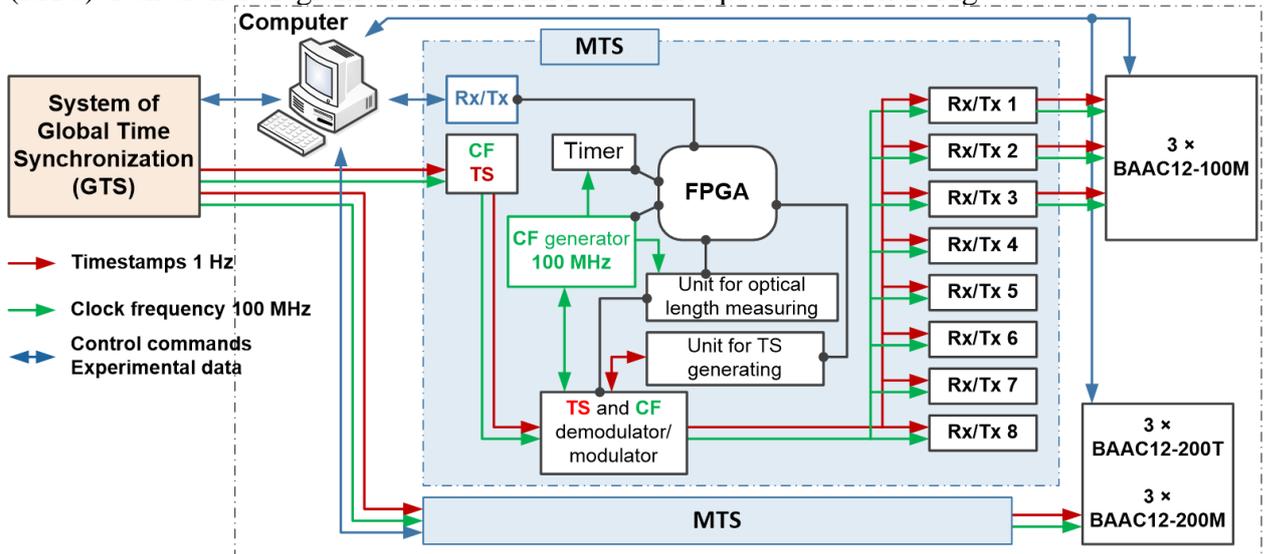

Fig. 4. Functional diagram of the PRISMA-36 central post.

The coordinated operation of the electronic units of the PRISMA-36 measuring system is ensured by the specialized software running on the control computer. For the data exchange between the MTS modules, BAAC12 blocks and the computer, the Ethernet network is used. Through the network interface, the computer manages the settings of the operating modes of the MTS modules and BAAC12 blocks and also receives information packets from BAAC12, containing waveforms of signals from its measuring channels and the time of event detection.

Upon receiving a BAAC12 packet, the control computer unpacks it, performs primary data processing and then records signal waveforms and results of their analysis.

The MTS module is based on the FPGA Xilinx Spartan-6 and contains eight optical communication channels, eight fiber-optic transceivers, one RJ-45 interface for communication with the control computer, one external synchronization channel and circuits for generating a 100 MHz clock frequency. This module can provide time synchronization of up to eight BAAC12 blocks. It distributes synchronization signals (timestamps (TS) with frequency of 1 Hz) and a byte (clock) frequency of 100 MHz (CF) to the connected BAAC12 blocks via fiber-optic lines. The BAAC12 blocks use TS and CF to clock their internal generators and timers. At this, the MTS modulates the clock frequency by the timestamp, and the BAAC12 modules demodulate this signal and restores the original CF and TS. This ensures synchronous operation of local timers and clock generators in the MTS and in the BAAC12 blocks connected to it.

Synchronization of the MTS modules with each other and, accordingly, the BAAC12 blocks connected to them is performed by the system of global time synchronization (GTS) of the EC NEVOD [23]. This system ensures the binding of events detected by the facilities of the complex to a single time source with an accuracy of 10 ns and to the world time with an accuracy of 1 s.

The variation channel of the PRISMA-36 array is implemented by reading out signal from the anode of the photomultiplier tube with its subsequent integration and amplification. To digitize signals, a block of amplitude analysis of a cluster with a sampling frequency of 100 MHz (BAAC12-100M) is used.

### 3. Photomultiplier tube

Scintillation flashes are detected using 8-inch 14-dynode hemispherical photomultiplier EMI 9350KA. This PMT has a bialkali photocathode. The maximal spectral sensitivity is achieved at an incident light wavelength of 360 nm. The quantum efficiency at such wavelength is 22–28% [24]. Overall dimensions and dynode system structure of the EMI 9350KA photomultiplier are shown in Fig. 5. These photomultipliers were previously used at the Gran Sasso laboratory in Italy [25, 26].

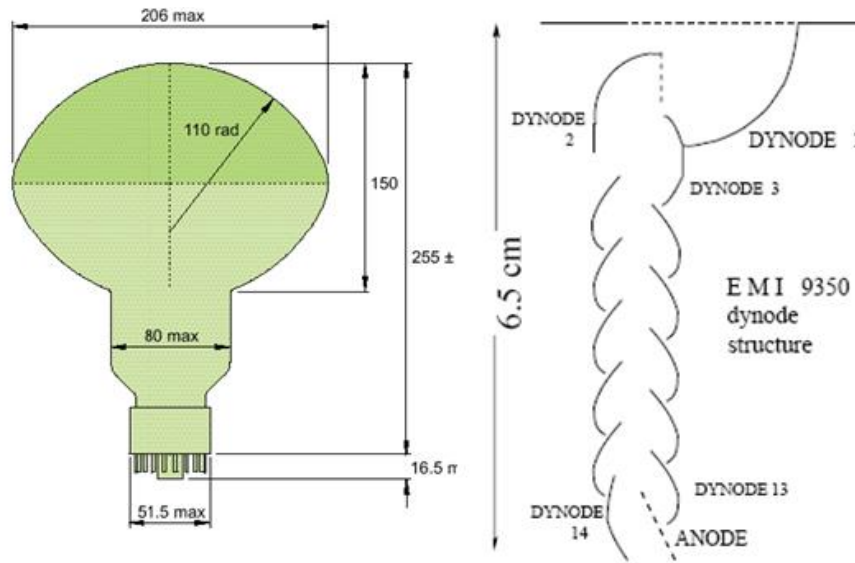

*Fig. 5. Overall dimensions (left) and dynode system structure (right) of the EMI 9350KA photomultiplier [24].*

Before using the EMI photomultiplier in the PRISMA-36 array, its dynode system gain and linearity range were studied. At this, a divider with an increase in the potential difference between the senior dynodes starting from the 8$^{th}$ one (Fig. 6) was used. As a photon source, the KingBright L-7113NBC light-emitting diodes (LEDs) with a wavelength of 445 nm were used. A Teflon diffuser [27] was used to diffuse photons from the light-emitting diodes.

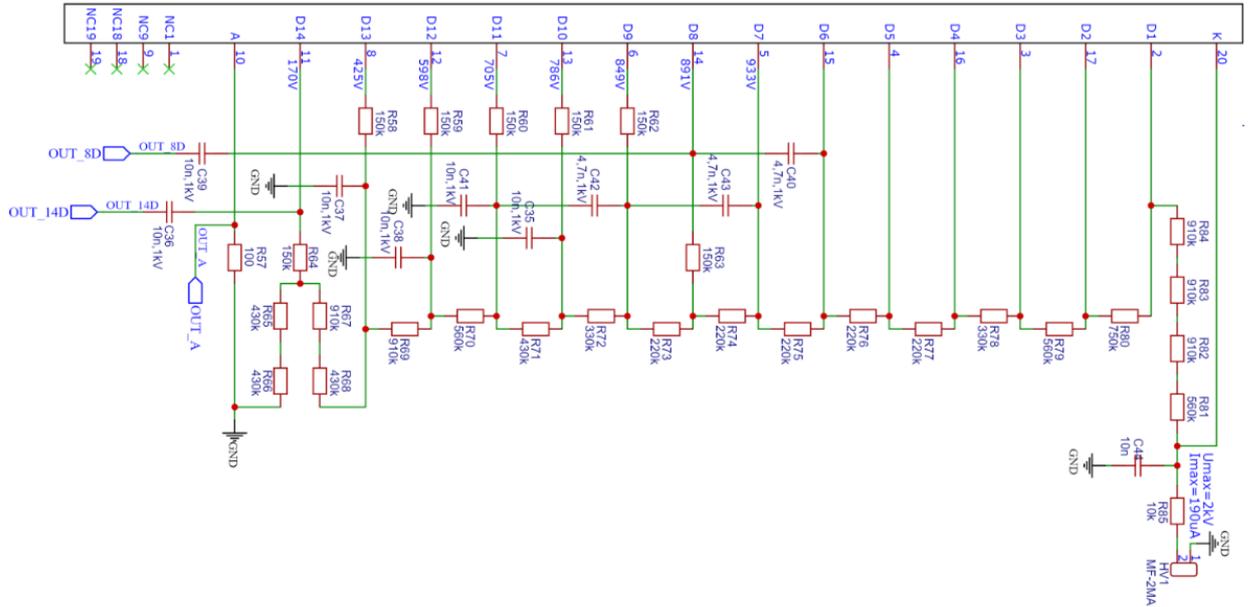

*Fig. 6. Basic circuit of the voltage divider for the EMI 9350KA photomultiplier.*

The gain of the dynode system was measured using the method of single-electron illumination, which is described in [28]. A typical charge distribution of anode signals from EMI 9350KA photomultiplier (No. 8436) is shown in Fig. 7. When testing the EMI 9350KA PMT No. 8436, a resolvable single-electron peak was obtained at a supply voltage of -1520 V. The valley is observed in the spectrum at a value of $Q_{val}$ = 0.18 pC. The peak-to-valley ratio is 2.3±0.1. The most probable value of the single-electron peak $Q_{peak}$ = 0.41 pC. The mean value and the standard deviation calculated for the part of the spectrum located to the right of the valley are $<Q>$ = 0.47±0.01 pC and $\sigma$ = 0.17±0.01 pC, respectively. The gain of the dynode system, estimated from the peak value [29], is $M$ = (2.55±0.01)×$10^6$.

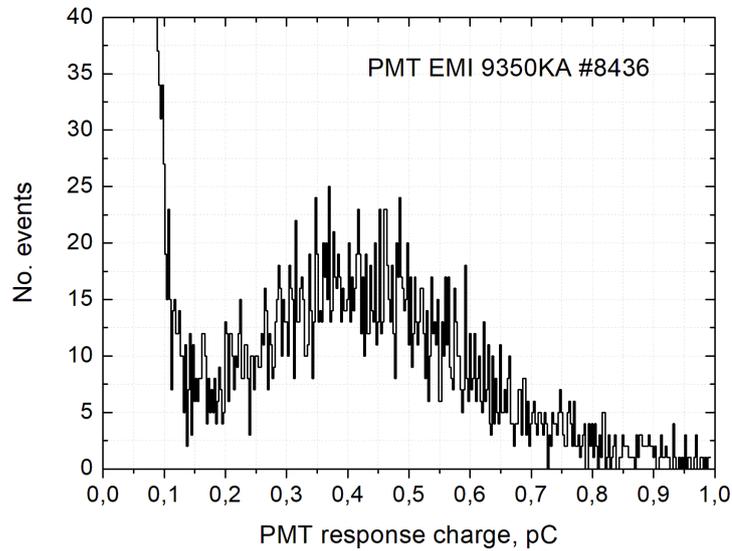

*Fig. 7. Typical distribution of charges of single-electron signals from the EMI 9350KA photomultiplier (No. 8436, supply voltage of -1520 V).*

Using the value M = 2.55×$10^6$ for the supply voltage of -1520 V, the dynode system gains were obtained in the supply voltage range from -1200 to -1680 V. For this purpose, the photocathode was illuminated by LED flashes with intensity ensuring the PMT response of 800 photoelectrons (ph.e.). Then the supply voltage of the PMT was reduced, and the measurements were repeated at the same LED brightness. For a known number of knocked-out photoelectrons, the gain was determined from the average charge of the anode signal. The dependence of the dynode system gain $M$ on the supply voltage $U$ is shown in Fig. 8. This dependence was approximated by the function:

$$M(U) = const\, U^w, \qquad (4)$$

and the exponent value $w = 11.7\pm0.3$ was obtained. This value is close to the expected exponent calculated using the formula:

$$w = N w_0, \qquad (5)$$

where $N$ is a number of dynodes, and $w_0$ is an empirical coefficient which is typically in the range from 0.6 to 0.8 [30]. In our case, the $w_0$ coefficient was 0.85.

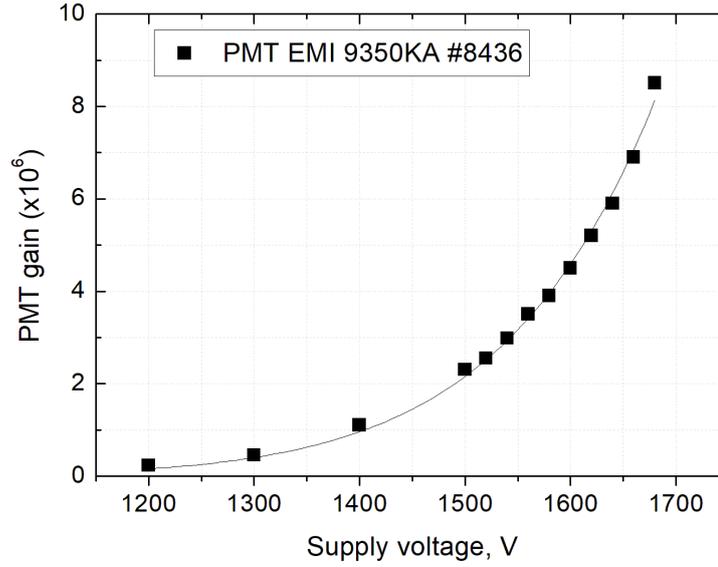

Fig. 8. Dependence of the dynode system gain on the supply voltage of the EMI 9350KA photomultiplier (No. 8436).

The obtained dependence made it possible to determine the operating voltage (-1360 V) which provides the PMT gain $M = 1\times10^6$.

The study of the PMT linearity range was carried out using the method of paired illumination. In this method, it is assumed that the PMT operates in the linear region, if its response to simultaneous illumination by two LEDs is approximately equal to the sum of the responses to illumination by each LED separately. For the numerical evaluation, the nonlinearity parameter $\alpha$ is used. It is calculated using the formula:

$$\alpha = \frac{Q_{1+2} - (Q_1 + Q_2)}{Q_1 + Q_2} \cdot 100\%, \qquad (6)$$

where $Q_{1+2}$ is the average charge of the anode signal, when the photocathode is illuminated by two LEDs simultaneously, $Q_1$ and $Q_2$ are the average charges of the anode signals, when PMT is illuminated by the both LEDs separately.

Average responses were calculated from the spectra with a statistical confidence of 5000 signals obtained for each measurement of $\alpha$ parameter. At each subsequent step, the intensity of the LED illumination was increased. It was considered, that the PMT operates in a linear mode, if the nonlinearity parameter does not exceed 5%. Fig. 9 shows the dependence of the nonlinearity parameter $\alpha$ on the charge of the output signal from the anode for the EMI 9350KA PMT No. 8436 with a supply voltage of -1360 V. For the divider circuit used and the dynode gain $M = 10^6$, the upper limit of the linearity range is about 700 pC or 4300 ph.e.

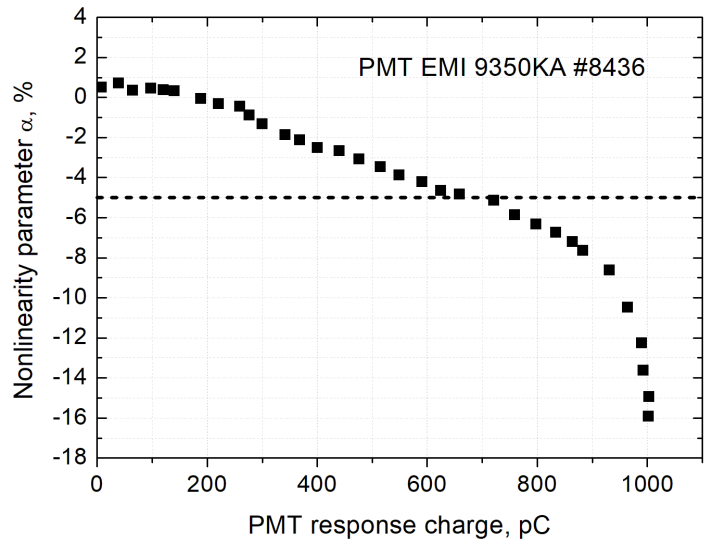

*Fig. 9. Dependence of the nonlinearity parameter α on the charge of the output signal from the anode of the PMT EMI 9350KA (No. 8436, supply voltage of -1360 V) under simultaneous illumination by two LEDs.*

Due to high quantum efficiency, large photocathode area and wide linearity range the EMI 9350KA PMT for the PRISMA-36 array are more promising for recording thermal neutrons in comparison with the FEU-200 photomultipliers previously used in the PRISMA-32 facility [31] and provide opportunity to increase the efficiency of neutron detection and neutron signals selection.

### 4. Integrating amplifier

To use the EMI PMTs in the PRISMA-36 array, a divider-integrator-amplifier (DIA) was developed. It provides independent detection of background neutrons, as well as the electron-photon and neutron components of extensive air showers. The DIA is a two-layer printed circuit board with a PMT high-voltage divider, three integrating amplifiers, power supply connectors (PMT high voltage in the range from 0 to -1900 V and low voltage ±12 V for the amplifiers) and connectors for reading out analog signals from the PMT anode (for studying neutron variations), as well as from the $14^{th}$ and $8^{th}$ dynodes of the photomultiplier (for EAS measurements).

The ZnS(Ag) scintillator has several components of luminescence from 0.1 to tens of μs when detecting α-particles and tritium nuclei, formed during the capture of neutrons by $^6$Li. Therefore, the analog integration of signals is used to increase the efficiency of detection and selection of neutron signals. The integration time is 2.7 μs for the anode signals and 1.5 μs for the dynode ones.

The basic circuit of the DIA integrating amplifier, based on the AD8065 operational amplifier, is shown in Fig. 10. To protect the amplifier input from the reverse polarity signals, the BAV99T116 diode assembly is used. The TI3121ID buffer amplifier is used to eliminate signal distortion during transmission over a cable line.

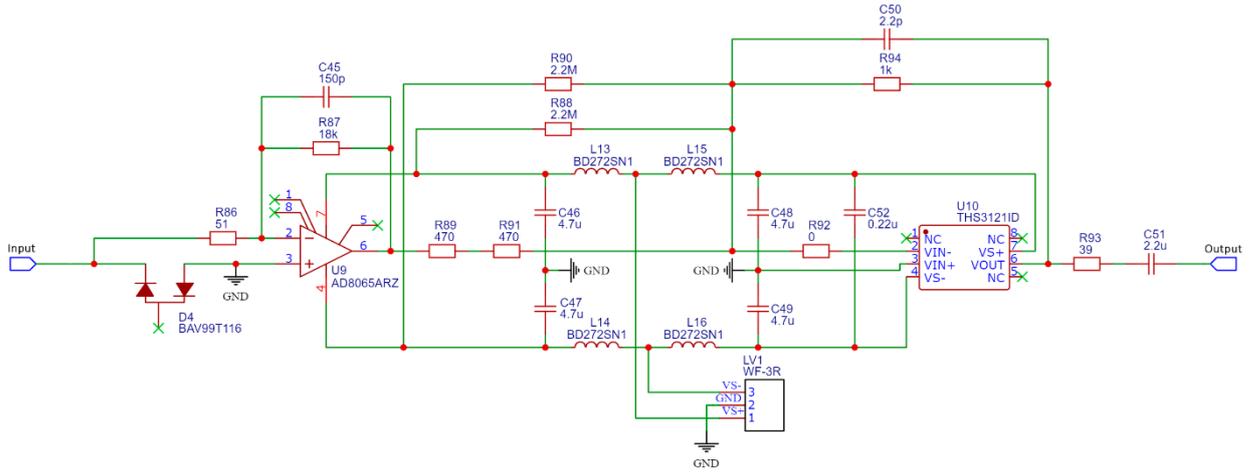

*Fig. 10. Basic circuit of the integrating amplifier of the DIA anode channel.*

Before using in the PRISMA-36 array, the conversion coefficients of the DIA integrating amplifiers for the neutron flux detection were measured. Positive polarity pulses (40 ns leading edge, 56 ns trailing edge, 400 ns duration) were fed from the arbitrary function generator Tektronix AFG3251 to the input of the corresponding channel. Output signals were read out with the digital oscilloscope Tektronix MDO3035, transferring signal waveforms to the computer for further processing.

A typical dependence of the amplitude of the signal at the amplifier output on the charge of the input signal is shown in Fig. 11. The coefficient of conversion of the input charge to the output voltage amplitude, obtained by approximating the experimental points with a linear function, is 4.00±0.01 mV/pC. The spread of the conversion coefficients of all 36 integrating amplifiers of the array is less than 3% from the average value given above. Such range allows converting PMT signals with charges from 0 to 1075 pC. Since the linearity range of the PMT anode channel is from 0 to 700 pC, the integrating amplifier does not limit the linearity of the photodetector signals.

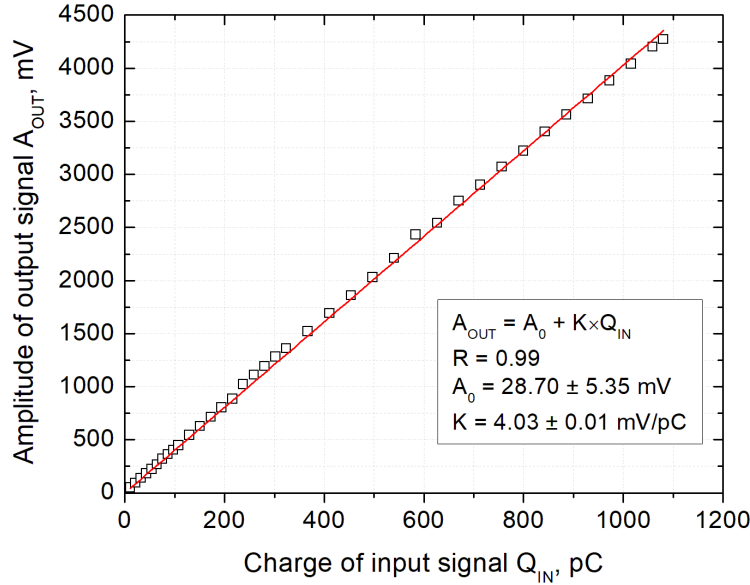

*Fig. 11. Dependence of the output signal amplitude on the charge applied to the input of the integrating amplifier.*

### 5. Block of amplitude analysis of a cluster BAAC12

The BAAC12-100M block digitizes analog signals from the variation channels of a cluster, selects events based on intra-cluster trigger conditions, performs timestamping of data and transmits them to the central post of the PRISMA-36 array (Fig. 12). Its main elements are:

- six dual-channel ADCs with a resolution of 12 bits and a sampling frequency of 100 MHz for digitalizing analog signals in the amplitude range from -3.5 to +3.5 V;

- a Xilinx Spartan-6 FPGA implementing various processor functions, i.e., reading out data from the ADCs, generating signal waveforms and their preliminary processing, selecting events based on trigger conditions, information exchange, etc.;
- a channel for synchronization with an external module of time synchronization via fiber-optic communication line;
- the circuits for generating clock frequency of 100 MHz for various functional elements;
- an interface (RJ-45) for receiving control commands and data transmitting.

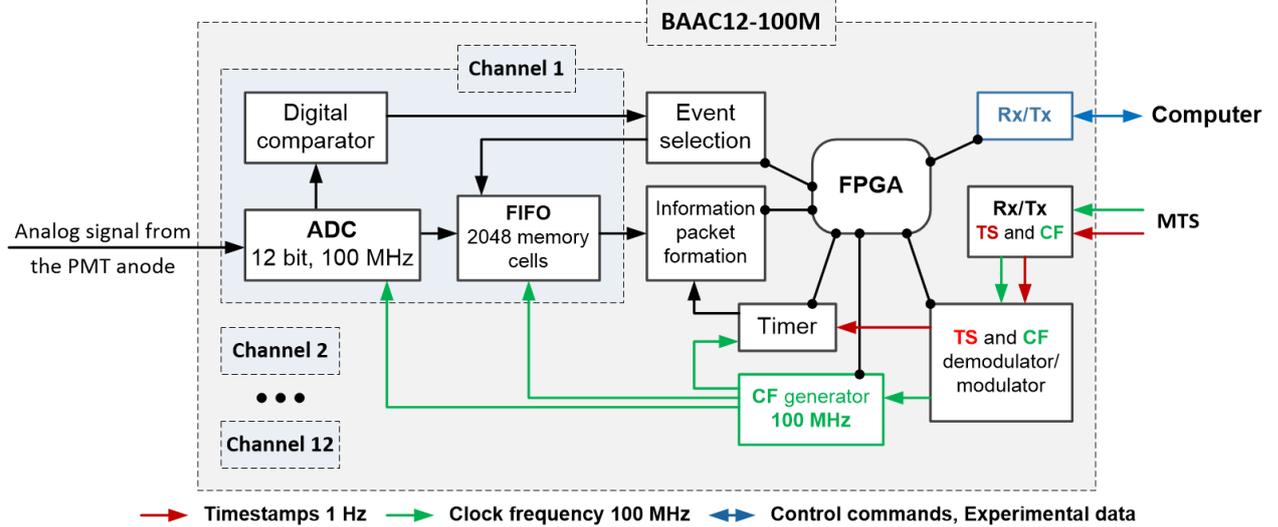

*Fig. 12. Functional diagram of the BAAC12-100M block.*

To study variations of the thermal neutron background, the BAAK12-100M blocks record PMT signals waveforms with a duration of about 20 μs (2048 samples of 10 ns each) on each of the 12 channels.

The BAAK12-100M blocks have internal clock generators operating at a frequency of 100 MHz and internal timers associated with them. The timers are used for timestamping of detected events with an accuracy of 10 ns. Synchronization of the internal clock generators and timers is carried out by the module of time synchronization (MTS), which is the part of the array central post.

During the calibration of the BAAC12-100M blocks, the ADC conversion coefficient and the linearity range were determined for each channel. To do this, a test facility consisting of a computer, a signal generator, a BAAC12-100M and a power supply was used. A rectangular pulse with a duration of 60 ns and an amplitude from 50 to 3550 mV with a step of 100 mV was fed to the input of each channel. A digital code corresponding to the input amplitude value was determined from the waveform read out from the BAAK12-100M. For each value of input signal amplitude, 1000 signals were recorded. The measured amplitude (in units of ADC lsb) spectrum was fitted by the Gauss distribution, and its peak value was determined. The dependence of the ratio between input and output signals on the output signal amplitude for 12 measuring channels of the BAAC12-100M block is show in Fig. 13.

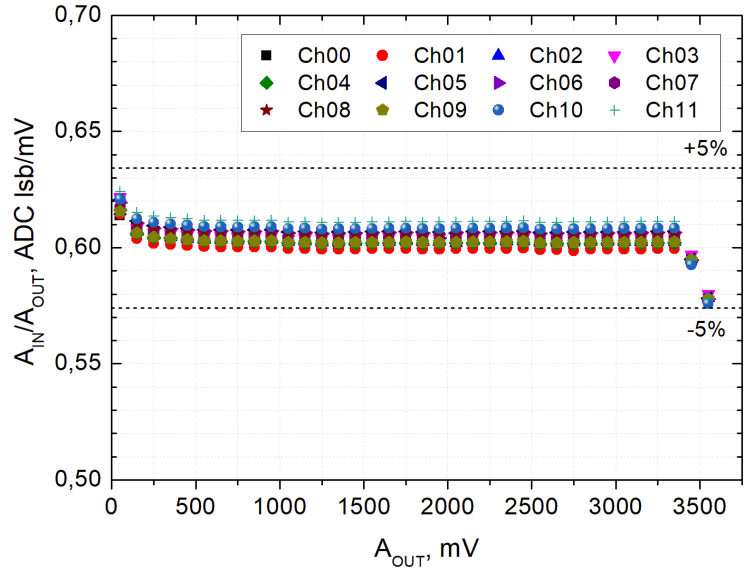

*Fig. 13. Dependence of the ratio between input and output signals on the output signal amplitude for 12 measuring channels of the BAAC12-100M block.*

From the obtained dependencies, the linearity range and the conversion coefficient (the ratio between amplitudes of the output and input signals) of the channels were determined. The average conversion coefficient <K> for 36 channels is 0.61±0.04 ADC lsb/mV. The ADCs linearity range, within which the absolute value of the ratio between amplitudes of the output and input signals do not exceed 5%, is from 0 to 3450 mV.

### 6. Detection of neutron signals

Since the used scintillator has different time components of luminescence when detecting charged particles and neutrons, it is possible to separate the signals by the pulse shape. For such scintillators, various methods of signal separation are known from the literature. In particular, among them are the use of analog or mathematical integration and separation of signals according to various ratios of their parameters [32–37], as well as approaches based on machine learning [38]. For the PRISMA-36 array, a method using analog integration was chosen. It facilitates signal transmission over long distances (cable length of 30 m) and noise suppression.

#### 6.1. Criteria for selection of neutron signals

Examples of integrated signals from neutrons and PMT noise are shown in Fig. 14.

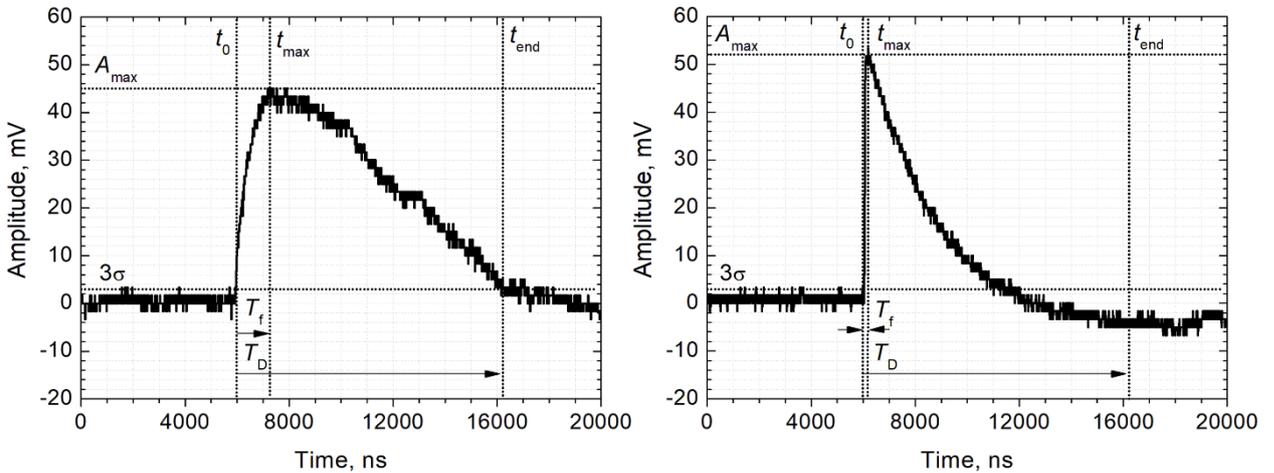

*Fig. 14. Examples of integrated neutron (left) and PMT noise (right) signals.*

It's seen, that the integrated signals from neutrons and PMT noise can have the same amplitude, but the neutron signals have a longer rise time and duration compared to the noise ones.

Based on the differences in these signals, the parameters for their separation (duration and front rise time) were selected.

When determining the parameters of signals, the value of the ADC pedestal and its standard deviation σ are calculated in the range from 0 to 3000 ns. Then the signal start $t_0$ and end $t_{end}$ times are obtained by searching the points in the waveforms, for which the corresponding conditions are met for the first time: $A(t_i) \geq 3\sigma$ and $A(t_i) \leq 3\sigma$, where $A(t_i)$ is the amplitude of the signal in the $i$-th waveform sample. The difference between these times $T_D = t_{end} - t_0$ is the signal duration. The signal front rise time $T_f$ is defined as the difference between the time, corresponding to the maximal amplitude $t_{max}$ and the start time $t_0$.

To determine the values of the parameters of neutron signals and PMT noise, two measurements were carried out: the first one was without a scintillator, the second one was with a scintillator and a neutron source. A certified neutron radiation source $^{252}$Cf (NCf2.82, neutron flux with energy of 2.12 MeV in 4π sr is ~ 9.6×10$^2$ s$^{-1}$, verified on 18.08.2023) was used. To thermalize fast neutrons to thermal energies, the source was placed in a polystyrene-based moderator with dimensions of 20×20×20 cm$^3$. The neutron source in the moderator was installed under the detector. The BAAC12-100M block was used to digitize the signals. Each signal with an amplitude greater than 8 ADC lsb (~13 mV) was saved.

To determine the criteria for selecting neutron signals, the dependences of the signal front rise time on the signal duration for measurement without a scintillator and without a source (Fig. 15a), as well as for one with a scintillator and a neutron source (Fig. 15b) were considered. Both measurements have the same duration of 3 hours.

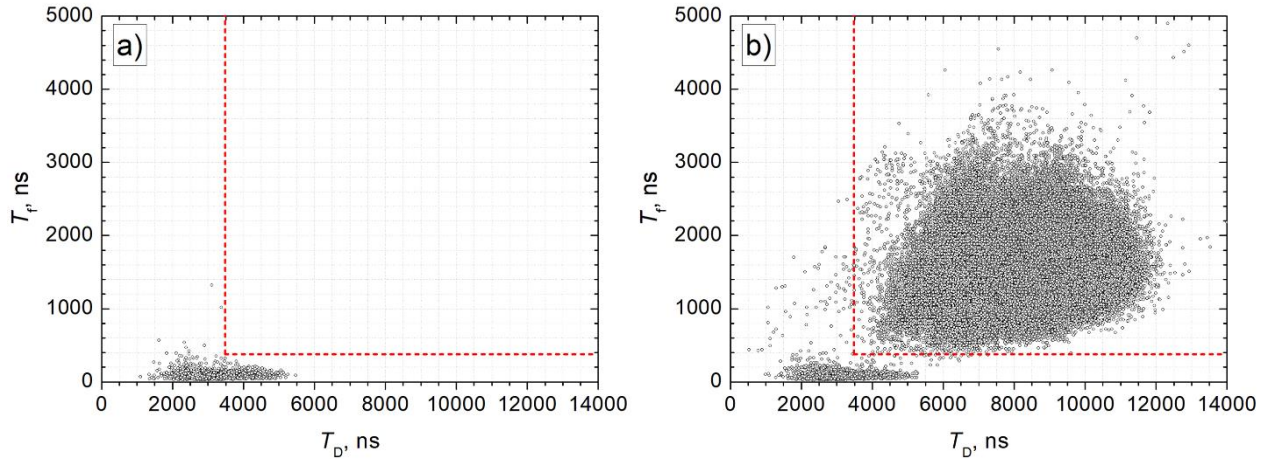

*Fig. 15. Dependences of the signal front rise time on the signal duration (a – without scintillator, b – with scintillator and neutron source $^{252}$Cf).*

When measuring without a scintillator (Fig. 15a), an area of events caused by photomultiplier noise pulses is observed. When installing a scintillator and a neutron source (Fig. 15b), another area related to neutron signals is observed in addition to events from PMT noise. The criteria for signal separation were determined in such a way that they would cover the area of neutron signals as much as possible, while signals from PMT noise would not be identified as neutrons. In the figures, the lines indicate the boundaries of the neutron signal area by the pulse front rise time ($T_f$=400 ns) and duration ($T_D$=3500 ns). Thus, the signal is identified as a neutron one when two conditions are met simultaneously: $T_f \geq 400$ ns and $T_D \geq 3500$ ns. Fig. 16 shows the dependence obtained from measurement of the neutron background, carried out with a scintillator, but without a neutron source. A similar dependence will be observed when detectors are operating as part of the installation.

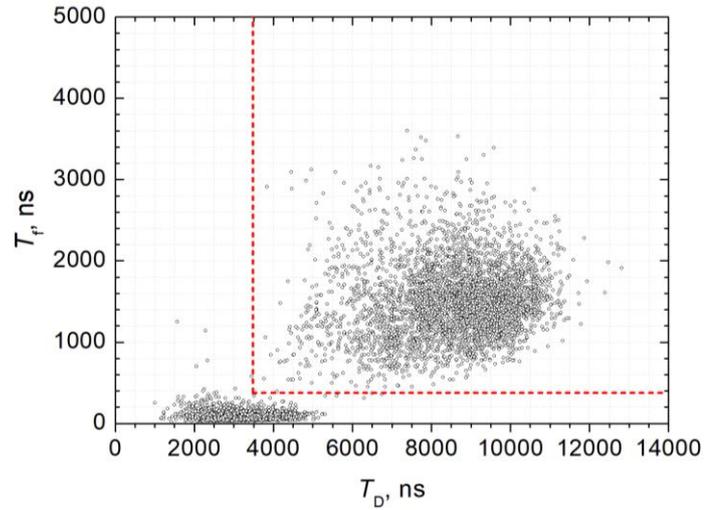

*Fig. 16. Dependence of the signal front rise time on the signal duration measured with the presence of scintillator, but without the neutron source $^{252}Cf$.*

Based on the data from the measurements described above, taking into account the defined criteria for signal separation, it is possible to obtain the counting rates of neutrons and noise signals (Fig. 17). In the period from 00:00 to 03:00 hours, the counting rates of neutron and noise signals in the absence of a scintillator are presented. During the measurement of natural background of neutrons carried out in the presence of a scintillator but without a source (from 03:00 to 06:00 hours), the averaged neutron counting rate was $0.35\pm0.03$ s$^{-1}$. When the source was installed, the neutron counting rate increased to $6.94\pm0.11$ s$^{-1}$. It can also be seen that the PMT noise counting rate did not change ($0.12 \pm 0.02$ s$^{-1}$), which indicates that the defined criteria are correct.

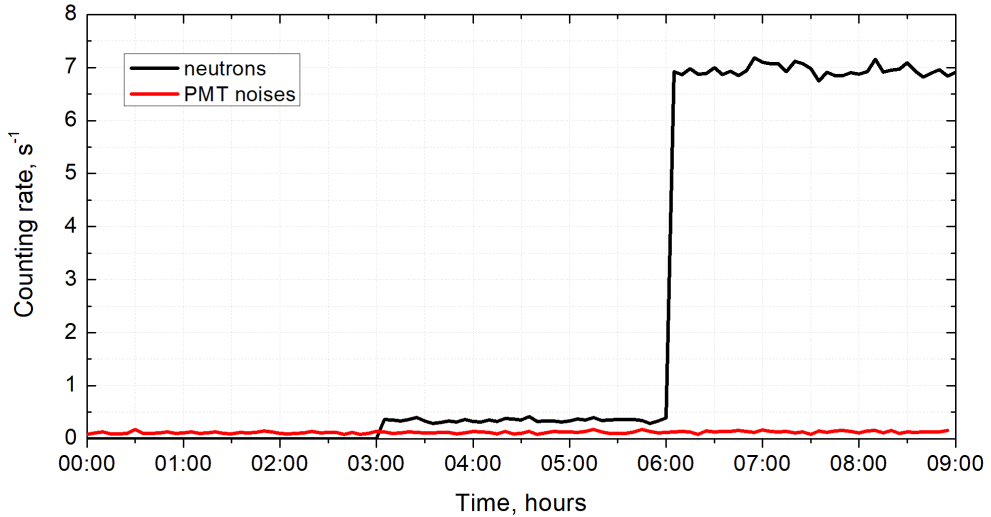

*Fig. 17. Counting rates of neutron and noise signals (at 03:00 the scintillator was installed, at 6:00 the neutron source $^{252}Cf$ inside the plastic moderator was added).*

**6.2. Efficiency of neutron detection**

Fig. 18 shows the distribution of the amplitudes of neutron signals selected taking into account the criteria described above. The distribution of neutron signal amplitudes can be described by an exponential function of the form $k \times e^{-bA}$. For this distribution, the *b* parameter is $0.045\pm0.001$ lsb$^{-1}$, and k is $0.071\pm0.006$. Having restored the distribution with accounting for the obtained exponent, it is possible to estimate the number of neutrons captured by the scintillator. The ratio of detected neutrons to the captured ones is $58.8\pm0.3\%$. Since the efficiency of thermal neutron capture by the SL6-5 scintillator is 20% [39], the efficiency of neutron detection with the chosen selection criteria is about 12%.

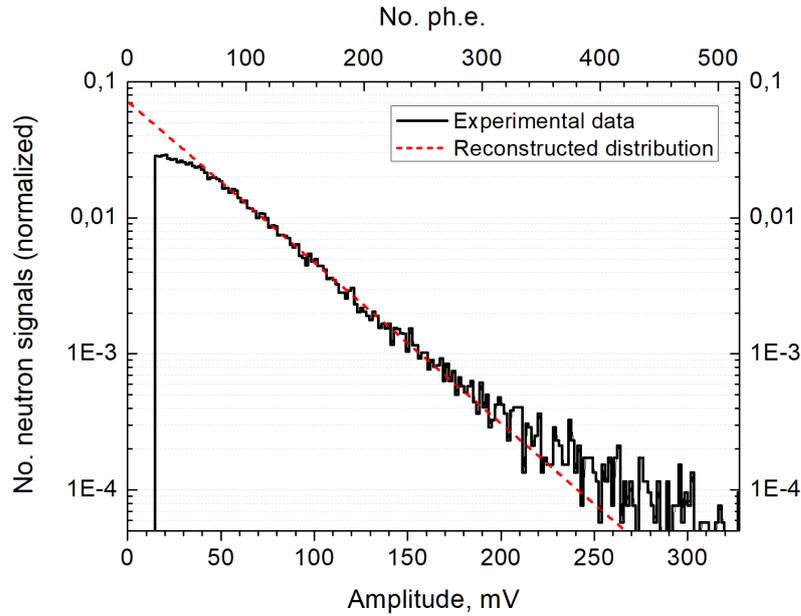

*Fig. 18. Distribution of neutron signal amplitudes normalized by their total number.*

Taking into account the gain of the PMT ($1.0 \times 10^6$) and the conversion factor of the spectrometric path (0.64 mV/ph.e.) and using the Fig. 18 (see the top axis), the number of photoelectrons, detected by the photomultiplier tube, per neutron capture can be determined. Thus, the threshold for selecting neutron signals is about 25 ph.e. Or it is 100 photons, if we take the PMT quantum sensitivity equal to 0.25 [24].

### 7. Results of test measurements

From October 2023 to March 2024, the test runs of data taking were carried out on the first cluster of the PRISMA-36 array. The trigger condition was the presence of a signal from at least one of 12 detectors with an amplitude greater than 13 mV. Such threshold value ensures a ratio between useful and noise signals from the detectors, optimal for the separation of thermal neutrons.

### 7.1. Measuring of neutron background

An example of detecting neutron and noise signals by three detectors of the PRISMA-36 is shown in Fig. 19. The average detectors counting rate is from 0.6 to 0.7 $s^{-1}$ for neutrons, and it is from 0.1 to 0.2 $s^{-1}$ for noise signals. Taking into account the detector area and the results of the detection efficiency estimation, the flux of background neutrons is $2 \times 10^{-3}$ $s^{-1} \cdot cm^{-2}$. The obtained value is consistent with the neutron radiation background near the Earth's surface [40].

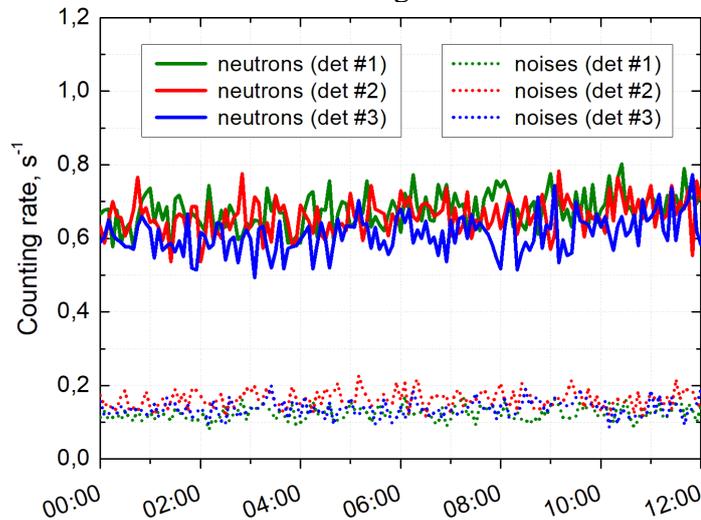

*Fig. 19. Counting rate of neutron and noise signals under background conditions for three detectors of the PRISMA-36 array.*

### 7.2. Barometric coefficient

Since the main part of the neutron background near the Earth's surface consists of particles, generated as a result of the interaction of cosmic rays with the nuclei of matter near the detectors, the background level should be influenced by the atmospheric pressure. The dependence of the total counting rate of 12 detectors on the atmospheric pressure, obtained using the data of the Vaisala weather station of the EC NEVOD [41], is shown in Fig. 20. A linear dependence of the counting rate of neutrons on the atmospheric pressure is observed. The barometric coefficient determined from the presented dependence is $\beta = (B/<N>) \times 100\% = -0.76 \pm 0.06\%$/mbar. It is consistent with the barometric coefficient for neutron monitors of the NM-64 type ($\beta_{MNM} = -0.723\%$/mbar) [42].

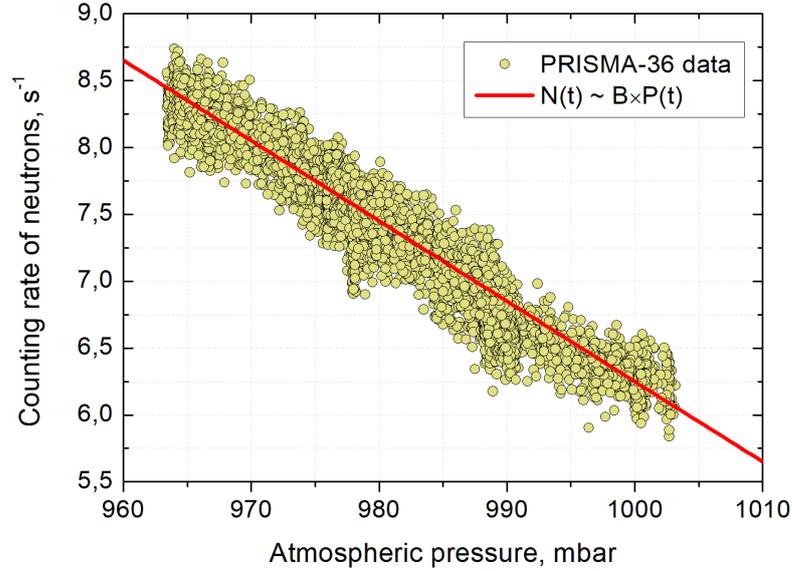

*Fig. 20. The dependence of the total counting rate of 12 detectors on the atmospheric pressure (December 2023).*

### 7.3. Response to Forbush decrease

Taking into account the correction for the barometric coefficient, the data on variation of the neutron flux, normalized to the average counting rate values, were obtained. The current counting rate value $N(t)$ was corrected according to the formula:

$$N_{corr}(t) = N(t) + B \cdot (P_0 - P(t)), \qquad (9)$$

where $P$ is the current atmospheric pressure, $P_0$ is the average atmospheric pressure, $B$ is the slope coefficient of the dependence $N(P)$ (-0.054±0.003).

Fig. 21 shows an example of neutron background variations for the period from March 23, 2024 to March 24, 2024, normalized to the average counting rate. During this period at 15:04 on March 24, a Forbush decrease was observed. For comparison, the data of the Moscow Neutron Monitor [43] are presented. According to the LASCO coronagraph [44], a coronal mass ejection (CME) with an average velocity of 942 km/s was recorded from an X1.1-class flare at 01:36 UTC on March 23.

Fig. 21 also shows the global disturbance of the Earth's magnetic field in a three-hour time interval according to OMNIWeb data [45]. The disturbance from the CME was detected on March 24. The peak disturbance was observed at 15:00 UTC. The maximal Kp×10 index of the global disturbance of the Earth's magnetic field in the mid-latitudes was 83 units.

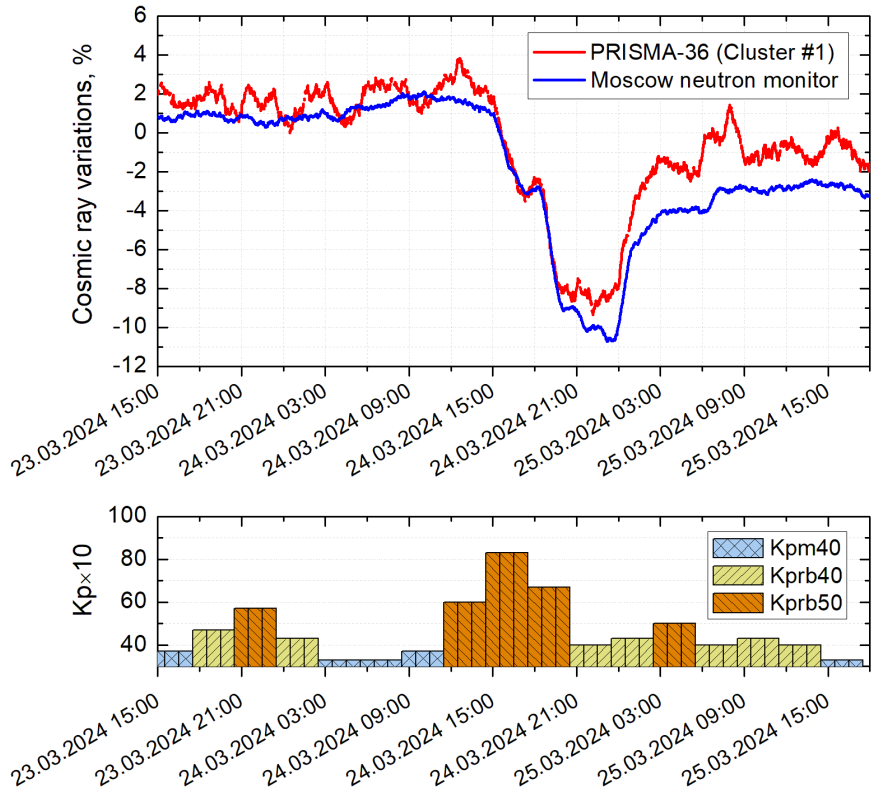

*Fig. 21. Neutron flux variations recorded during the Forbush decrease according to the data of the PRISMA-36 (12 detectors) and the Moscow Neutron Monitor (top), and global disturbance of the Earth's magnetic field in a three-hour time interval according to OMNIWeb data (Kp×10 index) (bottom).*

It is seen, that the counting rates of the PRISMA-36 detectors and the Moscow Neutron Monitor are in good agreement. During the Forbush decrease, a sharp drop of the neutron counting rate was observed both in the neutron monitor and in the PRISMA-36 cluster. For the presented Forbush decrease, the amplitude of the counting rate drop was determined using the method described in [46]. For the PRISMA-36, the amplitude of the drop was 12.3%±0.1%, and for the Moscow Neutron Monitor it was 11.8%±0.1%.

Thus, the unshielded neutron detectors based on ZnS(Ag) scintillator with $^6$LiF, which are the part of the PRISMA-36 array, are capable of measuring neutron flux variations with a sensitivity not inferior to a classical neutron monitor.

**8. Conclusion**

The PRISMA-36 array for studying variations in neutron fluxes, that are in thermal equilibrium with the environment, has been created. The array is based on unshielded neutron detectors with ZnS(Ag)+$^6$LiF scintillators and EMI 9350KA hemispherical photomultiplier tubes. It is deployed on an area of 500 m$^2$. The total area of the scintillators is about 13 m$^2$.

For the developed design of detector and detecting system, the detector response to thermal neutrons has been investigated, and the neutron signals selection criteria, ensuring rejection of PMT noise signals, has been determined. The efficiency of thermal neutron detection using these criteria for signal selection is estimated to be approximately 12%.

The PRISMA-36 variation channel for detecting thermal neutrons allows measuring flux values at the background level, as well as conducting studies of neutron flux variations with an accuracy not inferior to classical neutron monitors.

**Acknowledgements**

The work was performed at the Unique Scientific Facility "Experimental Complex NEVOD" with the support of the Russian Science Foundation (grant No. 23-22-00399,

https://rscf.ru/project/23-22-00399/). The authors express their gratitude to Yu.V. Stenkin (INR RAS), who participated in the development of the PRISMA project in MEPhI.

43. Data of the Moscow Neutron Monitor: http://cr0.izmiran.ru/mosc/ (accessed on October 14, 2024).
44. The CACTUS online CME catalog: http://sidc.oma.be/cactus/catalog.php (accessed on October 14, 2024).
45. The official web site of the OMNIWeb database: https://omniweb.gsfc.nasa.gov/form/dx1.html (accessed on October 14, 2024).
46. N.S. Barbashina, A.N. Dmitrieva, K.G. Kompaniets, A.A. Petrukhin, D.A. Timashkov et al., Specific Features of Studying Forbush Decreases in the Muon Flux, Bull. Russ. Acad. Sci. Phys. 73 (2009) 343-346.
19